\documentclass[aps,prd,reprint,balancelastpage,nofootinbib,preprintnumbers,amsmath,amssymb,superscriptaddress,longbibliography]{revtex4-1}
\usepackage[
colorlinks=true,        
citecolor=blue,         
linkcolor=blue,         
urlcolor=blue           
]{hyperref}             
\usepackage{bm}         
\usepackage{xcolor}     
\usepackage{orcidlink}
\usepackage{subfigure}
\newcommand{\nc}{\newcommand*}

\newcommand {\nn}{\nonumber}
\graphicspath{{Figures/}} 
\nc{\Eq}[1]{Eq.~\eqref{#1}}     
\nc{\Fig}[1]{Fig.~\ref{#1}}     
\nc{\Table}[1]{Table~\ref{#1}}  
\nc{\Sec}[1]{Sec.~\ref{#1}}     
\def\({\left(}
\def\){\right)}
\def\[{\left[}
\def\]{\right]}
\def\e{\begin{equation}}
\def\q{\end{equation}}
\def\m{\begin{eqnarray}}
\def\n{\end{eqnarray}}

\begin{document}

\title{Quasinormal modes of scalar perturbations in Rastall thick brane}

\author{Shan Huang\orcidlink{0000-0003-4936-5775}}
\email{hs_jcut@163.com}	
\affiliation{School of Mathematics and Physics, Jingchu University of Technology, Jingmen 448000, China}

\author{Chun-Chun Zhu\orcidlink{0000-0002-3389-1511}}
\email{zhucc@jcut.edu.cn, corresponding author}	
\affiliation{School of Mathematics and Physics, Jingchu University of Technology, Jingmen 448000, China}

\author{Tao-Tao Sui\orcidlink{0000-0003-3541-6842}}
\email{ suitaotao@aust.edu.cn, corresponding author}	
\affiliation{Center for Fundamental Physics, School of Mechanics and Photoelectric Physics, Anhui University of Science and Technology, Huainan 232001, Anhui, China}

\begin{abstract}
We investigate quasinormal modes of the graviscalar sector in a five-dimensional thick brane model in Rastall gravity. By considering a specific flat brane solution supported by a canonical scalar field, we derive a master equation and reduce it to a Schr\"odinger-like eigenvalue problem for the Kaluza-Klein modes. Using the Bernstein spectral method and direct integration in the frequency domain, complemented by numerical time-domain evolutions, we compute the complex quasinormal frequencies for the scalar perturbations. Our results reveal a strong dependence of the QNM spectrum on $\lambda$: the imaginary parts of the frequencies, governing the decay rate, decrease monotonically with increasing $\lambda$, indicating longer-lived modes. The real parts exhibit a more complex, non-monotonic behavior. Furthermore, we analyze the late-time behavior of the perturbations, showing that the asymptotic tail follows a power law whose exponent is determined by the Rastall parameter, in agreement with theoretical predictions for the asymptotic form of the potential. These findings provide a comprehensive dynamical characterization of the scalar sector of Rastall thick branes, offering potential observational signatures for probing modified gravity in extra-dimensional scenarios.
\end{abstract}
\maketitle

\section{Introduction}
\label{sec:intro}

The braneworld picture assumes that our observable universe is a four-dimensional hypersurface embedded in a higher-dimensional spacetime. This idea appears naturally in Kaluza--Klein (KK) and string-inspired models~\cite{kaluza:1921un,Klein:1926tv}, and it also provides new approaches to the hierarchy problem~\cite{Arkani-Hamed:1998jmv,Antoniadis:1998ig,Randall:1999ee,Randall:1999vf}. A key feature of the RS-II model is that four-dimensional gravity can be effectively reproduced on the brane even when the extra dimension is noncompact. This happens because the graviton zero mode can be localized by the warped geometry.

Thick brane models replace the idealized thin brane by smooth field configurations in the bulk, usually supported by scalar fields coupled to gravity~\cite{DeWolfe:1999cp,Gremm:1999pj,Csaki:2000fc}. Compared with thin branes, thick branes have richer internal structure. This structure changes the spectrum of KK excitations and affects the way perturbations propagate into the extra dimension. Therefore, thick branes provide a useful arena to study dynamical signals of extra dimensions beyond the static localization of zero modes~\cite{Afonso:2007gc,Dzhunushaliev:2010fqo,Dzhunushaliev:2011mm,Geng:2015kvs,Melfo2006,Almeida2009,Zhao2010,Chumbes2011,Liu2011,Bazeia:2013uva,Xie2017,Gu2017,ZhongYuan2017,ZhongYuan2017b,Zhou2018,Hendi:2020qkk,Xie:2021ayr,Moreira:2021uod,Xu:2022ori,Silva:2022pfd,Xu:2022gth,Tan:2020sys,Tan:2022uex,Zhu:2023tzx,Dzhunushaliev:2009va,Maartens:2010ar,Liu:2017gcn,Ahluwalia:2022ttu}.

A natural tool to probe such dynamics is quasinormal modes (QNMs). QNMs are damped oscillations of open systems. They are defined by purely outgoing boundary conditions at spatial infinities, and they form a discrete set of complex frequencies. In black hole physics, QNMs encode geometric information and are directly linked to ringdown signals in gravitational-wave observations~\cite{Berti:2009kk,Kokkotas:1999bd,Nollert:1999ji,Konoplya:2011qq,Cardoso:2016rao,Jusufi:2020odz,Cheung:2021bol,Chen:2024gwf,Liu:2024iso,Guo:2022rms,Guo:2023nkd}. In braneworld systems, QNMs have a different but equally clear meaning. They describe massive KK waves that stay near the brane for a finite time and then leak into the bulk. For an observer on the brane, these modes behave like unstable massive particles. Because the spectrum depends on the effective potential along the extra dimension, QNMs can serve as a ``spectroscopy'' of the brane structure~\cite{Seahra:2005wk,Seahra:2005iq,Tan:2022vfe,Tan:2023cra,Jia:2024pdk,Tan:2024url,Tan:2024aym,Tan:2024dbl,Jia:2024sdk,Deng:2025hfn,Zhu:2024gvl,E:2025kic}.

Most previous thick-brane QNM studies concentrate on transverse-traceless tensor perturbations, i.e., the spin-2 sector. This is well motivated, since tensor modes are directly related to gravitational waves and the recovery of Newtonian gravity. However, tensor perturbations are only part of the full fluctuation content of a scalar-field thick brane. After a scalar-vector-tensor decomposition, the metric fluctuations split into tensor, vector, and scalar sectors~\cite{Giovannini:2001fh,Giovannini:2001xg,Giovannini:2001vt,Kobayashi:2001jd}. 
The scalar sector is often called the graviscalar sector. It is tightly connected to the bulk scalar configuration that generates the brane. This sector is absent in the original RS thin-brane setup without matter sources, but it is unavoidable in thick branes. 
Therefore, studying graviscalar modes is important if one wants to use dynamical signals to distinguish thick branes from thin branes.

The graviscalar sector also has a phenomenological reason to be studied carefully. A localized scalar zero mode would mediate an extra long-range force on the brane, which is strongly constrained. In many thick brane models the scalar zero mode is non-normalizable and thus not localized, which avoids this problem. Even in this case, the massive graviscalar KK continuum can still produce characteristic ringing. Recent analyses in general relativity indicate that graviscalar perturbations admit discrete QNMs and show clear late-time tails, including power-law tails for effectively massless evolution and oscillatory tails for massive evolution~\cite{Tan:2024aym,Deng:2025hfn}. These late-time behaviors depend on the asymptotic form of the effective potential and thus provide another window into the geometry of the extra dimension.

In this work we consider thick branes in Rastall gravity. Rastall gravity modifies the usual conservation law of the energy-momentum tensor in curved spacetime by allowing a nonzero divergence proportional to the curvature gradient~\cite{Rastall:1972swe,Harko:2014gwa}. It can be viewed as an effective description of nonminimal matter-gravity coupling, and it has been widely explored in cosmology and compact objects~\cite{BezerradeMello:2014okn,Heydarzade:2017wxu,Darabi:2017tay,Xu:2017vse,Darabi:2017coc,Das:2018dzp,Tang:2019dsk,Li:2019jkv,Khyllep:2019odd,Ghosh:2021byh,Haghani:2022lsk,Shahidi:2021lxt,Vagnozzi:2022moj}. For braneworlds, a thick brane solution in Rastall gravity was constructed in Ref.~\cite{Zhong:2022wlw}. A striking result is that the tensor sector differs from the standard general-relativity case: the flat Rastall thick brane does not support a normalizable graviton zero mode for nonzero Rastall parameter. This motivates the study of metastable and quasinormal behavior, because quasi-localized modes may still mimic four-dimensional gravity within a finite range~\cite{Tan:2024url}. While the tensor QNM spectrum has been investigated, a systematic QNM analysis of the scalar gravitational sector in the Rastall thick brane is still lacking. {The master variable of the scalar sector mixes the scalar-field fluctuation with the scalar part of the metric, its effective potential is not the tensor potential of Ref.~\cite{Tan:2024url}, and its asymptotic $1/z^2$ coefficient generates a $\lambda$-dependent tail exponent. These features determine whether the scalar sector contains dangerous long-range components and how it rings, which cannot be inferred from the spin-2 spectrum.} This is precisely the goal of the present paper.

We focus on the graviscalar perturbations of the Rastall thick brane and compute their quasinormal spectrum. We derive the master equation in the longitudinal gauge and reduce it to a Schr\"odinger-like eigenvalue problem for the KK modes. We then compute the quasinormal frequencies in the frequency domain using several complementary methods, including the direct integration and the Bernstein spectral method. To verify and interpret the spectrum, we also evolve initial wave packets in the time domain, extract dominant frequencies by fitting, and study the late-time tails. We pay special attention to how the Rastall parameter changes the effective potential and, in turn, the QNM frequencies and tail exponents. These results provide a coherent dynamical picture of the scalar gravitational sector of Rastall thick branes and extend the current QNM program in warped extra dimensions.

This paper is organized as follows. In Sec.~\ref{sec:model}, we review the background thick brane solution in Rastall gravity and derive the master equation for scalar perturbations. In Sec.~\ref{sec:qnms}, we compute the quasinormal frequencies in the frequency domain and confirm them in the time domain. We analyze the late-time tails based on the asymptotic structure of the effective potential. Finally, we conclude in Sec.~\ref{sec:conclusion}.

\section{Thick brane and scalar perturbations}
\label{sec:model}

\subsection{Background solution in Rastall gravity}

In Rastall gravity, the nonconservation law is assumed as~\cite{Rastall:1972swe}
\begin{equation}
	\nabla^M T_{MN}=\lambda \nabla_N R,
	\label{eq:rastall}
\end{equation}
where $\lambda$ is the Rastall parameter. 
The five-dimensional field equation can be written as~\cite{Zhong:2022wlw}
{\begin{equation}
R_{MN}-\left(\frac12-\lambda\right)Rg_{MN}=\kappa_5T_{MN},
\label{eq:field}
\end{equation}
where $\kappa_5$ is the five-dimensional fundamental scale. Bulk indices are $M,N=0,1,2,3,5$, and brane indices are $\mu,\nu=0,1,2,3$. For simplicity in calculation, we will set $\kappa_5=1$.} Observations suggest $|\lambda|\ll 1$~\cite{Akarsu:2020yqa}. 

\begin{figure}
	\subfigure[~the warp factor~\eqref{eq:warp}]{\label{warpfactor1}
		\includegraphics[width=0.23\textwidth]{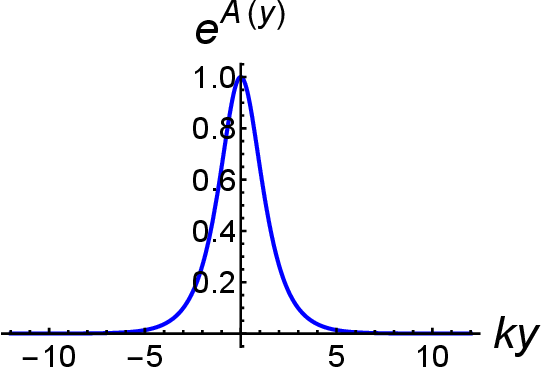}}
	\subfigure[~the scalar field~\eqref{eq:scalar}]{\label{scalarfield1}
		\includegraphics[width=0.23\textwidth]{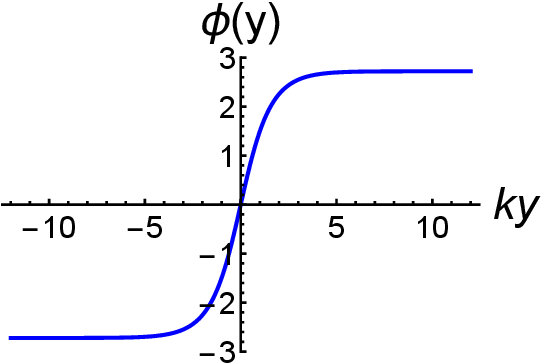}}
	\subfigure[~the scalar potential~\eqref{eq:potential}]{\label{scalarpotential1}
		\includegraphics[width=0.23\textwidth]{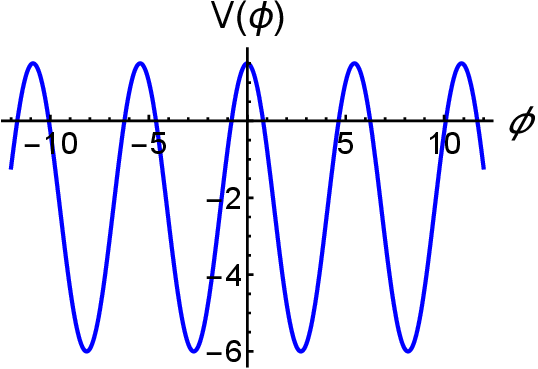}}
	\caption{Plots of the warp factor~\eqref{eq:warp}, the scalar field~\eqref{eq:scalar}, and the scalar potential~\eqref{eq:potential}.}\label{scalarandwarpfactor}
\end{figure}

We consider a canonical scalar field $\phi$ with
\begin{equation}
	T_{MN}=\partial_M\phi\,\partial_N\phi-\frac12 g_{MN}\left(\partial^A\phi\,\partial_A\phi+2V(\phi)\right).
\end{equation}
The background metric in conformal coordinates is
\begin{equation}
	ds^2=e^{2A(z)}\left(\eta_{\mu\nu}dx^\mu dx^\nu+dz^2\right),
	\label{eq:metric}
\end{equation}
where $\eta_{\mu\nu}=\mathrm{diag}(-1,1,1,1)$. 
Substituting this ansatz into the field equations gives
\begin{align}
	3A'^2-3A''-\phi'^2&=0, \label{eq:dyn1}\\
	2e^{2A}V(\phi)+3(3+8\lambda)A'^2+(3+16\lambda)A''&=0, \label{eq:dyn2}
\end{align}
where primes denote derivatives with respect to $z$. 
A simple thick-brane solution is~\cite{Zhong:2022wlw}
\begin{align}
	A(z)&=-\frac12\ln(1+k^2 z^2), \label{eq:warp}\\
	\phi(z)&=\sqrt{3}\arctan(kz), \label{eq:scalar}\\
	V(\phi)&=\frac{k^2}{4}\Big[(56\lambda+15)\cos\!\left(\frac{2\phi}{\sqrt{3}}\right)-3(8\lambda+3)\Big],
	\label{eq:potential}
\end{align}
with $k$ a mass scale. The shapes of warp factor, scalar field, and scalar potential are shown in Fig.~\ref{scalarandwarpfactor}.  Note that Eqs.~\eqref{eq:dyn1} and \eqref{eq:dyn2} are identical to the corresponding equation in general relativity, so $A(z)$ and $\phi(z)$ keep the same form.  The parameter $\lambda$ only enters through the effective potential.

\subsection{Scalar perturbations and the master equation}

We study scalar perturbations in the longitudinal gauge~\cite{Kobayashi:2001jd}:
\begin{equation}
	ds^2=e^{2A(z)}\Big[(1+2\varphi)\eta_{\mu\nu}dx^\mu dx^\nu+(1+2\Psi)dz^2\Big],
	\label{eq:perturbed}
\end{equation}
and $\phi\to \phi(z)+\delta\phi(x^\mu,z)$. 
Linearizing the field equations, the off-diagonal $(\mu\neq\nu)$ components give the constraint
\begin{equation}
	\varphi+2\Psi=0.
	\label{eq:constraint1}
\end{equation}
From the $(z,\mu)$ components, we can express $\delta\phi$ in terms of $\Psi$ and $\varphi$:
\begin{equation}
\delta\phi=\frac{-3\partial_z\Psi+3A'\varphi}{\phi'}.
\label{eq:constraint2}
\end{equation}
Using Eqs.~\eqref{eq:constraint1} and~\eqref{eq:constraint2} to eliminate $\varphi$ and $\delta\phi$, we obtain a single master equation for $\Psi$:
\begin{eqnarray}
	\frac{10\lambda-3}{4\lambda-3}\Box^{(4)}\Psi+\left(3A'-\frac{2\phi''}{\phi'}\right)\partial_z\Psi
	+\partial_z^2\Psi\nn\\
	+\left(\frac{4(44\lambda-3)}{4\lambda-3}A'^2+\frac{4(3-20\lambda)}{12\lambda-9}{\phi'^2}-\frac{4A'\phi''}{\phi'}\right)\Psi
	=0,
	\label{eq:master}
\end{eqnarray}
where $\Box^{(4)}=\eta^{\alpha\beta}\partial_\alpha\partial_\beta$.

We now introduce a KK-type decomposition adapted to Eq.~\eqref{eq:master}:
\begin{equation}
	\Psi(x^\mu,z)=e^{-\frac32 A(z)}\,\phi'(z)\,\widetilde{\Psi}(t,z)\,e^{-i a_i x^i},
	\label{eq:KK}
\end{equation}
where $a=\sqrt{a^i a_i}$ is the spatial momentum along the brane. 
Then $\widetilde{\Psi}(t,z)$ satisfies
\begin{equation}
	-\partial_t^2 \widetilde{\Psi}+\partial_z^2\widetilde{\Psi}-U(z)\widetilde{\Psi}-a^2\widetilde{\Psi}=0.
	\label{eq:wave}
\end{equation}
Writing $\widetilde{\Psi}=e^{-i\omega t}\psi(z)$, we arrive at a Schr\"odinger-like equation
\begin{equation}
	-\psi''(z)+U(z)\psi(z)=m^2\psi(z),
	\label{eq:schrodinger}
\end{equation}
with $m^2=\omega^2-a^2$.

\begin{figure}[htbp]
	\centering
	\includegraphics[width=0.45\textwidth]{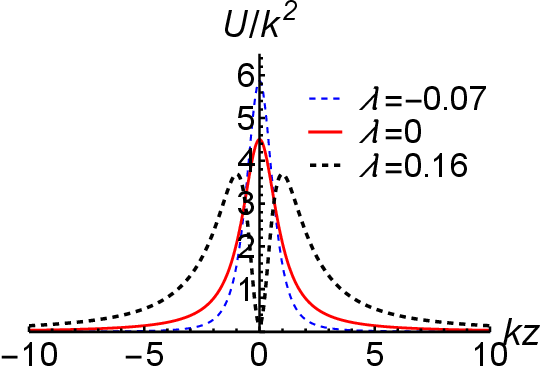}
	\caption{The shape of the graviscalar effective potential $U(z)/k^2$ as a function of $kz$.}
	\label{fig:potential}
\end{figure}

For the Rastall thick brane, the effective potential $U(z)$ can be written in a compact form:
\begin{eqnarray}
	U(z)&=&\frac{(3-644\lambda)A'^2}{4(4\lambda-3)}
	+\frac{A'\phi''-\phi^{(3)}}{\phi'}\nn\\
	&&+\frac{(148\lambda-15)\phi'^2}{6(4\lambda-3)}
	+\frac{2\phi''^2}{\phi'^2}.
	\label{eq:Ugeneral}
\end{eqnarray}
Substituting Eqs.~\eqref{eq:warp} and~\eqref{eq:scalar}, we obtain an explicit closed expression:
\begin{equation}
	U(z)=\frac{k^2\Big[328\lambda-5(116\lambda+9)k^2 z^2-54\Big]}{4(4\lambda-3)(1+k^2 z^2)^2}.
	\label{eq:Us}
\end{equation}
The shape of this potential is shown in Fig.~\ref{fig:potential}. As the parameter $\lambda$ increases, the effective potential becomes lower and broader. Since $U(z)\to 0$ as $|z|\to\infty$, the massive KK modes are scattering states and can leak into the bulk. {The scalar zero mode is not normalizable in the present model, so the relevant scalar excitations are massive scattering or QNM modes rather than a localized long-range scalar mode.} The sign of $U(z)$ depends on $\lambda$. In the range
\begin{equation}
	{-\frac{9}{116}\leq \lambda\leq\frac{54}{328},}
	\label{eq:lambdarange}
\end{equation}
the potential \eqref{eq:Us} is non-negative for all $z$, which is a convenient stable window for QNM computations. 
Outside this window, $U(z)$ can take negative values. This does not automatically imply an instability, but it makes tachyonic modes possible and requires a separate spectral check. 
{In this non-negative-potential window, the associated self-adjoint operator is non-negative, so normalizable scalar bound states and tachyonic bound states are absent. Once the potential develops negative regions, unstable modes with $\mathrm{Im}(m)>0$ may appear, so the stability has to be checked explicitly.}
{The two end points of \eqref{eq:lambdarange} have a simple interpretation for the explicit potential \eqref{eq:Us}. At $\lambda=-9/116$ the coefficient of the asymptotic $1/z^2$ tail vanishes, so the late-time power-law index reaches the edge of the Ching classification used below. At $\lambda=54/328$ the central value $U(0)$ vanishes. Crossing either value changes the sign pattern of the potential but does not introduce a singularity in the QNM boundary-value problem. Outside this range, the effective potential may take negative values and unstable modes may occur.}

\section{Quasinormal Modes}
\label{sec:qnms}

For QNMs we impose purely outgoing waves at both spatial infinities. 
Since $U(\pm\infty)=0$, Eq.~\eqref{eq:schrodinger} reduces asymptotically to $\psi''+m^2\psi\simeq 0$. 
We choose the convention
\begin{equation}
	\psi(z)\propto 
	\begin{cases}
		e^{+im z}, & z\to +\infty,\\[2mm]
		e^{-im z}, & z\to -\infty,
	\end{cases}
	\label{eq:BC}
\end{equation}
which corresponds to waves traveling away from the brane on both sides. 
With this boundary condition, the allowed $m$ form a discrete set of complex values. 
For stable modes, $\mathrm{Im}(m)<0$, so the time dependence $e^{-i\omega t}$ gives a decaying signal at fixed $z$.
\subsection{Frequency Domain}
\label{sec:frequency}
In this section, we compute the graviscalar quasinormal frequencies using two methods: the direct integration method~\cite{Pani:2013pma} and the Bernstein spectral method~\cite{Fortuna:2020obg}. Since the Bernstein spectral method has the highest accuracy, we mainly use it to solve the QNMs. Other methods serve as evidence.

The Bernstein spectral method~\cite{Fortuna:2020obg} is a powerful numerical technique for solving eigenvalue problems by expanding the solution in a basis of Bernstein polynomials. For a general linear differential operator $\hat{L}(u, \omega)$:
\begin{equation}
	\hat{L}(u, \omega) \Phi(u) = 0, \quad u \in [a, b],
\end{equation}
the solution is expanded as:
\begin{equation}
	\Phi(u) = \sum_{k=0}^{N} C_k B_k^N(u),
	\label{eq:Bernstein}
\end{equation}
where $B_k^N(u) = \binom{N}{k} \frac{(u-a)^k (b-u)^{N-k}}{(b-a)^N}$ are Bernstein polynomials. Substituting this expansion leads to a generalized eigenvalue problem for the coefficients $C_k$, which can be solved using standard numerical linear algebra techniques.

\begin{figure}
	\subfigure[~The first QNM]{\label{figlambdarem1}
		\includegraphics[width=0.22\textwidth]{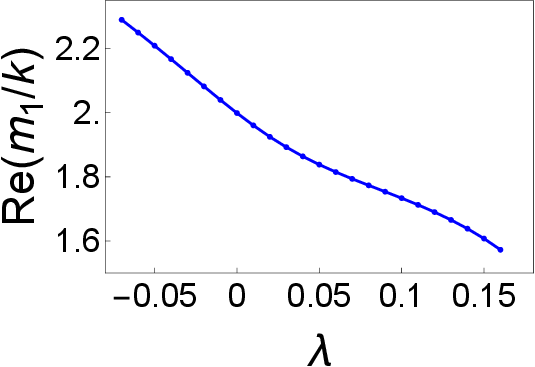}}
	\subfigure[~The first QNM]{\label{figlambdaimm1}
		\includegraphics[width=0.22\textwidth]{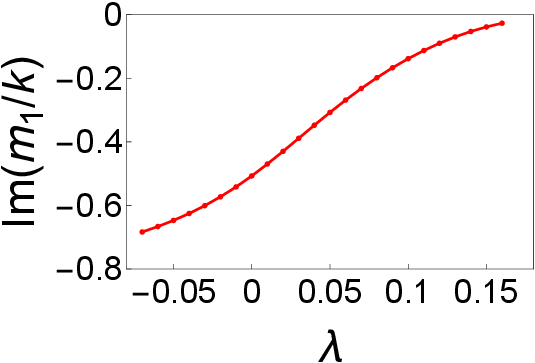}}
	\subfigure[~The second QNM]{\label{figlambdarem2}
		\includegraphics[width=0.22\textwidth]{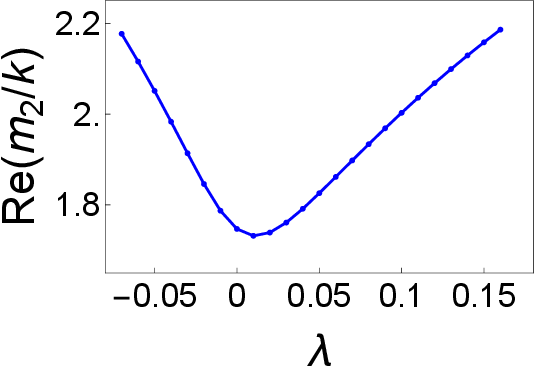}}
	\subfigure[~The second QNM]{\label{figlambdaimm2}
		\includegraphics[width=0.22\textwidth]{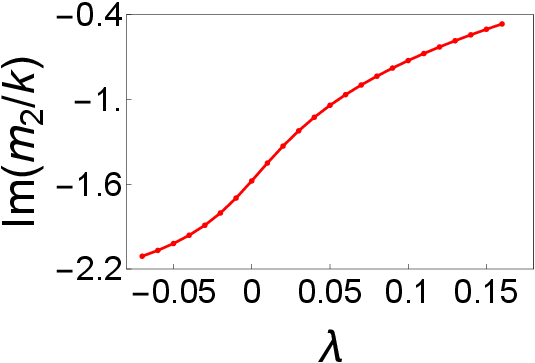}}
	\subfigure[~The third QNM]{\label{figlambdarem3}
		\includegraphics[width=0.22\textwidth]{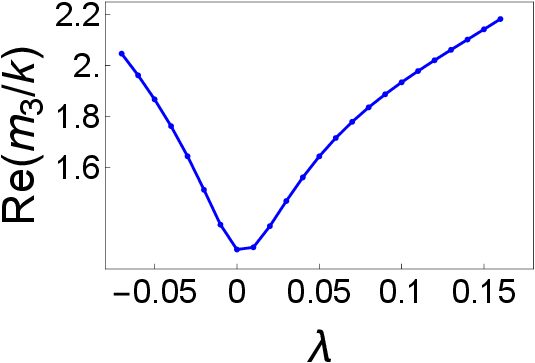}}
	\subfigure[~The third QNM]{\label{figlambdaimm3}
		\includegraphics[width=0.22\textwidth]{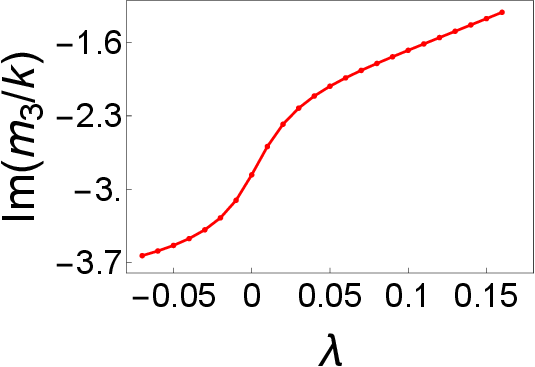}}
	\caption{Left panel: The relation between the real parts of the first three QNMs and $\lambda$. Right panel: The relation between the imaginary parts of the first three QNMs and $\lambda$.}\label{lambdagt0}
\end{figure}

After compactifying the infinite domain $z \in (-\infty, +\infty)$ to $u \in [-1, 1]$ via the coordinate transformation~$u= \frac{\sqrt{4k^2 z^2+1}-1}{2k z}$ and incorporating the outgoing wave boundary conditions, we apply the Bernstein spectral method to compute the quasinormal frequencies. The results are shown in Table~\ref{tab1} and Fig.~\ref{lambdagt0}. The real part of the fundamental QNM and the absolute values of the imaginary parts of the first three QNMs decrease monotonically with the Rastall parameter $\lambda$, indicating that the corresponding lifetimes increase. By contrast, the real parts of the two higher overtones first decrease and then increase with $\lambda$, forming a V-shaped trend, as shown in Figs.~\ref{figlambdarem2} and~\ref{figlambdarem3}. {This non-monotonicity has a simple physical origin. Increasing $\lambda$ lowers the central height of the potential but also broadens the effective scattering region. For the higher overtones, which have shorter wavelengths and are more sensitive to the detailed phase accumulated across the barrier, these two effects compete: the lower barrier tends to reduce the oscillation frequency, whereas the broader trapping region increases the phase path. The minimum of $\mathrm{Re}(m)$ marks the point where these effects balance.} Furthermore, as shown in Table~\ref{tab1}, the quasinormal frequencies obtained by the frequency-domain methods and the resolvable time-domain fits are in good agreement, which supports the reliability of our results.

\begin{table*}[htbp]
	\begin{tabular}{|c|c|c|c|c|}
		\hline
		$\;\;\lambda\;\;$  &
		$\;\;n\;\;$  &
		$\;\;\text{Bernstein spectral method}\;\;$  &
		$\;\;\;\;\;\;\;\;\text{Direct integration method}\;\;\;\;\;\;\;$ 	&
		$\;\;\;\;\;\;\;\;\text{Time evolution}\;\;\;\;\;\;\;$\\
		\hline
		~  &~   &~~~~$\text{Re}(m/k)$  ~~  $\text{Im}(m/k)~~$  &$~~~~~~\text{Re}(m/k)$ ~~ $\text{Im}(m/k)~~$  &$~~~~~~\text{Re}(m/k)$ ~~ $\text{Im}(m/k)~~$     \\
		{54/328}  &{1}   &{1.55445~~ -0.022030}          &{~~1.55445~~ -0.022030}           &{~~1.55445~~ -0.0220298}     \\
        &{2}   &{2.19887~~ -0.448101}          &{~~2.19887~~ -0.448101}                &{~~2.19884~~ -0.448062} \\
		0.16  &1   &1.57237~~ -0.026550          &~~1.57237~~ -0.026549                &~~1.572374~~ -0.026549     \\
		&2   &2.18625~~ -0.464986          &~~2.18625~~ -0.464986                &~~2.186209~~ -0.464925 \\
		0.07  &1   &1.79374~~ -0.232122          &~~1.79374~~ -0.232122                &~~1.793743~~ -0.231970     \\
		&2   &1.89787~~ -0.897285          &~~1.89787~~ -0.897285                &~~1.898705~~ -0.906472 \\		
		0.04  &1   &1.86395~~ -0.347754          &~~1.86395~~ -0.347754                &~~1.863949~~ -0.347742  \\
		&2   &1.79140~~ -1.126180          &~~1.791403~~ -1.126180                &{~~--~~} \\
		0.01~~ &1   &1.96065~~ -0.469812          &~~1.96065~~~-0.469812                &~~1.961208~~~-0.470709 \\
		&2   &1.73181~~ -1.450330          &~~1.73203~~~-1.450267                &{~~--~~} \\
		0~~&1  &1.99929~~ -0.507338         &~~1.99929~~  -0.507338                &~~1.999394~~  -0.503601 \\
		-0.01 &1   &2.04014~~ -0.541866          &~~2.04014~~ -0.541866    &~~2.045823~~   -0.545288\\	
		&2   &1.78713~~ -1.698030          &~~1.79713~~  -1.60658               &{~~--~~} \\	
		-0.04 &1   &2.16762~~ -0.625474        &~~2.16762~~ -0.625474    &~~2.177626~~ -0.625823 \\
		&2   &1.98335~~ -1.961800          &~~1.98811~~ -1.970202                &{~~--~~}  \\
		-0.07~~&1   &2.29044~~ -0.683709          &~~2.29044~~ -0.683709  &~~2.234926~~~ -0.689133 \\
		&2   &2.17715~~ -2.110120          &~~2.10155~~ -2.100979                &{~~--~~} \\
		{-9/116}~~&{1}   &{2.31988~~ -0.695475}          &{~~2.31988 ~~ -0.695475}  &{~~2.279503~~~ -0.699205} \\
		&{2}   &{2.22122~~ -2.137732}          &{~~2.20652~~ -2.122549}                &{~~--~~} \\
		\hline
	\end{tabular}
	\caption{The first two QNMs using Bernstein spectral method, the direct integration method, and the time evolution. {The symbol ``--'' indicates modes not robustly extractable from the time-domain signal.} \label{tab1}}
\end{table*}

{We further check representative points outside the conservative interval \eqref{eq:lambdarange}. For comparison with the time-domain evolution discussed below, the frequency-domain roots and the time-domain fitted values are collected together in Table~\ref{tab:outside}. The table shows that the unstable scalar modes appear for sufficiently negative $\lambda$ and for sufficiently large positive $\lambda$. At the same time, the case $\lambda=0.2$ has $\mathrm{Im}(m)<0$, providing an explicit example that a partially negative potential does not necessarily imply an instability. Thus the non-negative-potential interval is a sufficient, but not necessary, stability criterion.}

\begin{table*}[htbp]
	\begin{tabular}{|c|c|c|c|}
		\hline
		$\lambda$ &
		$\text{Bernstein spectral method}$ &
		$\text{Time evolution}$ &
		$\text{behavior}$ \\
		\hline
		~ & $\text{Re}(m/k)\quad \text{Im}(m/k)$ & $\text{Re}(m/k)\quad \text{Im}(m/k)$ & ~\\
		\hline
		{ $-0.1$} & { $0\quad\quad 0.018608$} & { $1.04071\times10^{-9}\quad 0.018571$} & { unstable} \\
		{ $-0.2$} & { $0\quad\quad 0.329751$} & { $0.000838\quad\quad 0.331043$} & { unstable} \\
		{ $-0.3$} & { $0\quad\quad 0.567648$} & { $0.000409\quad\quad 0.569903$} & { unstable} \\
		{ $0.2$} & { $1.36989\quad -0.003115$} & { $1.36989\quad\quad -0.003115$} & { decaying} \\
		{ $0.4$} & { $0~\quad\quad 2.51130$} & { $3.21634\times10^{-9}\quad 2.51129$} & { unstable} \\
		{ $0.6$} & { $0~\quad\quad 6.63245$} & { $7.33166\times10^{-8}\quad 6.63245$} & { unstable} \\
		\hline
	\end{tabular}
	\caption{Representative modes outside the non-negative-potential interval \eqref{eq:lambdarange}. A positive imaginary part of $m/k$ corresponds to an exponentially growing mode in time. {Regarding the time evolution, the listed small values ($1.04\times10^{-9}$, $3.22\times10^{-9}$, $7.33\times10^{-8}$) of  $\text{Re}(m/k)$ for $\lambda=(-0.1, 0.4, 0.6)$ represent the numerical floors for the fitting, which are consistent with $\text{Re}(m/k) = 0$.} \label{tab:outside}}
\end{table*}

\subsection{Time Domain}
\label{subsec:time}

To complement the frequency-domain analysis and study the dynamical behavior of graviscalar perturbations, we perform numerical time evolution of initial wave packets. Using light-cone coordinates $u = t - z$ and $v = t + z$, the wave equation~(\ref{eq:wave}) becomes:
\begin{equation}
	\left( 4\frac{\partial^2}{\partial u \partial v} + U + a^2 \right) \tilde{\Psi}(u, v) = 0.
	\label{eq:lightcone}
\end{equation}

\begin{figure}
		\subfigure[~$\lambda=-0.07$]{\label{figlogevenqnmsn7100}
	 \includegraphics[width=0.23\textwidth]{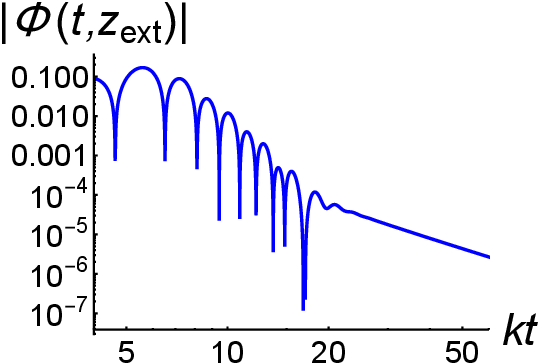}}
	 \subfigure[~$\lambda=-0.01$]{\label{figlogevenqnmsn1100}
	 \includegraphics[width=0.23\textwidth]{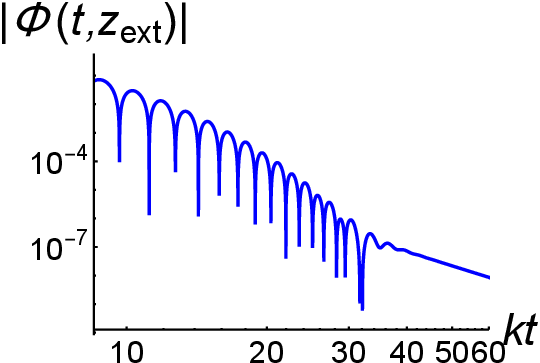}}
	 	 \subfigure[~$\lambda=0.07$]{\label{figlogevenqnms7100}
	 	 \includegraphics[width=0.23\textwidth]{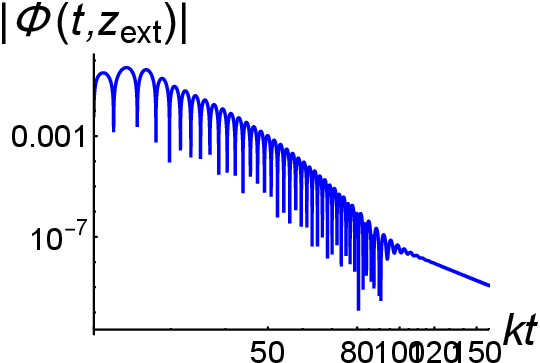}}
	 	 	 \subfigure[~$\lambda=0.16$]{\label{figlogevenqnms16100}
	 \includegraphics[width=0.23\textwidth]{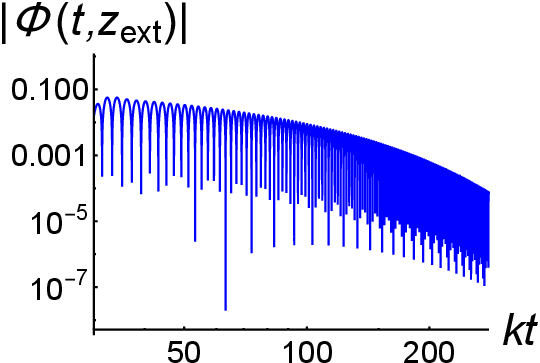}}
	\caption{Time evolution of Gaussian wave packet at $kz_{\text{ext}} = 3$ with different $\lambda$, plotted on a logarithmic scale.}
	\label{fig:evolution_a0}
\end{figure}

We discretize this equation using a finite difference scheme and evolve the initial data numerically. We consider a Gaussian wave packet
\begin{eqnarray}
	\tilde{\Psi}(0, v)&=& \exp\left( -\frac{(kv - kv_c)^2}{2k^2\sigma^2} \right), \nn \\
	\tilde{\Psi}(u, 0) &=& \exp\left( -\frac{k^2 v_c^2}{2k^2\sigma^2} \right),
\end{eqnarray}
with $kv_c = 5$ and $k\sigma = 1$. This initial packet excites all modes but is dominated by the fundamental mode. In this work we restrict to the case $a=0$. The waveform extracted at $kz_{\text{ext}} = 3$ is shown in Fig.~\ref{fig:evolution_a0}. It displays the three standard stages of QNM evolution: the initial burst, quasinormal ringing, and late-time tail. The waveform lifetime also increases with the Rastall parameter $\lambda$. The quasinormal frequencies extracted from the numerical evolution by fitting the waveform are presented in Table~\ref{tab1}, showing good agreement with the frequency-domain results. {The entries marked by ``--'' in Table~\ref{tab1} are not missing frequency-domain modes. They are higher overtones whose damping rates are so large that only a few oscillations survive after the initial burst. In the time signal they are rapidly masked by the fundamental mode and by the onset of the late-time tail, so a multi-mode Prony or nonlinear least-squares fit is not stable under changes of the fitting window. We therefore quote only time-domain frequencies that are robust against such changes.} The late-time tail also depends on the Rastall parameter $\lambda$, as discussed below.

\begin{figure}[htbp]
	\subfigure[~$\lambda=-0.1$]{\label{figevenqnmsn1f10}
		\includegraphics[width=0.23\textwidth]{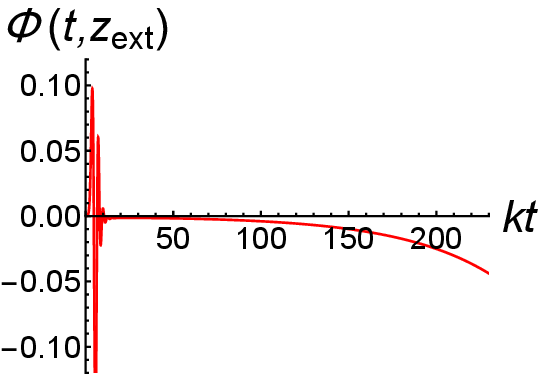}}
	\subfigure[~$\lambda=0.4$]{\label{figevenqnms4f10}
		\includegraphics[width=0.23\textwidth]{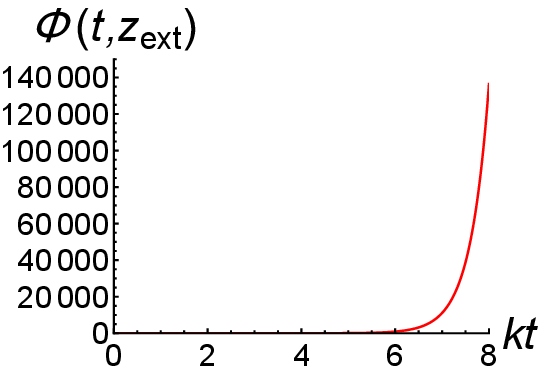}}
	\caption{{Time-domain evolution outside the non-negative-potential interval. The two examples show exponentially growing scalar modes, in agreement with the positive imaginary parts in Table~\ref{tab:outside}.}}
	\label{fig:outsideevolution}
\end{figure}

{We also evolved the representative outside-window points listed in Table~\ref{tab:outside}. For the unstable cases, the positive imaginary part obtained in the frequency domain is visible directly as exponential growth in the time-domain profiles. Two examples are shown in Fig.~\ref{fig:outsideevolution}.}

The late-time behavior of perturbations provides important information about the asymptotic structure of the effective potential. Following the analysis of Ching et al.~\cite{Ching:1994bd,Ching:1995tj}, for an effective potential with the asymptotic form:
\begin{equation}
	V(x) \sim \frac{\nu(\nu+1)}{x^2} + \frac{c_1 \ln x + c_2}{x^\alpha}, \quad x \to \infty,
\end{equation}
the late-time tail behavior depends on the parameter $\nu$.

When $c_1 = 0$:
\begin{itemize}
	\item If $\nu$ is an integer, the tail is $\tilde{\Psi} \sim t^{-\beta}$ with $\beta > 2\nu + \alpha$.
	\item If $\nu$ is not an integer, the tail is $\tilde{\Psi} \sim t^{-\beta}= t^{-(2\nu+2)}$.
\end{itemize}

For the graviscalar effective potential~(\ref{eq:Us}), the asymptotic behavior is:
\begin{equation}
	U(z) \sim \frac{5(116\lambda+9)}{4(3-4\lambda) z^2 }+ O(z^{-4}),
\end{equation}
corresponding to $\nu = 2\sqrt{\frac{ 36 \lambda +3}{3-4 \lambda }}-\frac{1}{2}$. We calculated the results under different $\lambda$ values and compared them with the tail behavior obtained from the numerical evolution fitting. The results are presented in Table~\ref{tab2}. It can be seen that the results obtained from the theory and numerical calculations are consistent.

\begin{table}[htbp]
	\begin{tabular}{|c|c|c|}
		\hline
		$\;\;\lambda\;\;$  &
		$\;\;\text{theoretical result}\;\;$  &
		$\;\;\;\;\;\;\;\;\text{fitting result}\;\;\;\;\;\;\;$ \\
		\hline
		~   &~  $\beta~~$  &~~~~ $\beta~~$       \\
		-0.07   &~~ 2.53018      &~~~ 2.57562       \\
		-0.01   &~~4.72756      &~~ 4.78344  \\
		0.07   &~~ 6.69830      &~~~~ 6.76007  \\
		\hline
	\end{tabular}
	\caption{The tail parameter $\beta$ with different values of $\lambda$ and different initial data calculated by fitting the late time data in Fig.~\ref{fig:evolution_a0}.\label{tab2}}
\end{table}

{We now give an order-of-magnitude interpretation of the above dimensionless spectrum, following the same dimensional strategy used in thick-brane QNM phenomenology~\cite{Tan:2022vfe,Tan:2024dbl}. Using natural units and taking the brane scale as $k=10^{-3}\,{\rm eV}$, the physical frequency and damping time of the $a=0$ modes are
\begin{equation}
	\nu_n=\frac{k}{2\pi}\mathrm{Re}\left(\frac{m_n}{k}\right),\qquad
	\tau_n=\frac{1}{k\left|\mathrm{Im}\left(m_n/k\right)\right|},
	\label{eq:physicalunits}
\end{equation}
where $n$ labels the QNM branch. In the same natural units, we quote the light-travel distance during one damping time as an upper scale for how far such a transient can effectively influence the brane,
\begin{equation}
	d_n\lesssim \tau_n=\frac{1}{k\left|\mathrm{Im}(m_n/k)\right|}.
	\label{eq:lifetimedistance}
\end{equation}
Laboratory tests of the inverse-square law strongly constrain any additional long-range scalar force~\cite{Adelberger:2003zx,Lee:2020zjt}. In the parameter window mainly considered in this work, the scalar zero mode is not normalizable, so the dangerous massless scalar channel is absent. The remaining modes in Table~\ref{tab1} are massive QNMs with finite lifetimes. For the fundamental modes in the non-negative-potential window, the representative value $k=10^{-3}\,{\rm eV}$ gives
\begin{equation}
	\begin{aligned}
	\nu_n&\sim 10^{11}\,{\rm Hz},\\
	\tau_n&\sim 10^{-12}-10^{-11}\,{\rm s},
	\end{aligned}
\end{equation}
and
\begin{equation}
	\begin{aligned}
	d_n&\lesssim 10^{-2}\,{\rm m}.
	\end{aligned}
\end{equation}
Thus the scalar signal is both short-lived and short-distance. These scalar QNMs therefore do not mediate a macroscopic fifth force. Their frequencies are also far above the LISA and pulsar-timing frequency bands, closer to the ultra-high-frequency regime discussed for KK-graviton signals in thick-brane models~\cite{Tan:2022vfe,Tan:2024dbl}.}

{A further phenomenological difference from the tensor-sector QNMs is the polarization content. The present perturbation is a graviscalar fluctuation, so a gravitational-wave interpretation would not correspond to the usual transverse-traceless tensor polarizations. It would instead require sensitivity to scalar-type polarizations, analogous to a breathing component in four-dimensional language. Current detector responses and bounds are usually optimized for tensor modes; a direct observational constraint on the present scalar spectrum would therefore require both a source model for the excitation amplitude and a detector or detector network capable of separating scalar polarizations. The main robust statement from the present calculation is consequently spectral: the Rastall parameter controls the lifetime and the late-time tail shape of any excited scalar signal.}

\section{Conclusions and Discussion}
\label{sec:conclusion}
In this work, we have systematically analyzed the quasinormal modes arising from graviscalar perturbations in a thick brane model embedded in five-dimensional Rastall gravity, complementing previous studies that focused primarily on the tensor sector. Starting from the background solution, we derived the master equation for the scalar perturbation and transformed it into a one-dimensional Schr\"odinger-like equation with an effective potential $U(z)$ that depends explicitly on the Rastall parameter $\lambda$. This potential vanishes asymptotically, allowing for a continuous spectrum of Kaluza-Klein modes, some of which manifest as quasinormal modes under outgoing wave boundary conditions.

Using a combination of frequency-domain methods (Bernstein spectral method and direct integration) and time-domain evolutions of Gaussian wave packets, we computed the complex quasinormal frequencies for the scalar perturbations. Our results reveal a strong sensitivity to the Rastall parameter: for the fundamental mode, increasing $\lambda$ decreases both the oscillation frequency and the damping rate, leading to longer-lived modes. The higher overtones exhibit more intricate behavior, with their real parts showing non-monotonic dependence on $\lambda$. {This behavior reflects the competition between the decreasing height and increasing width of the scalar effective potential, a feature absent from a purely tensor-sector interpretation.} The excellent agreement between independent numerical methods validates the robustness of our findings.

The late-time behavior of the perturbations follows a power-law decay, $\tilde{\Psi} \sim t^{-\beta}$, whose exponent is determined by the asymptotic form of the effective potential. We derived the analytic relation between $\beta$ and $\lambda$ from the potential's $1/z^2$ tail and confirmed it through numerical fitting, demonstrating consistency between theory and simulation. This connection highlights how the asymptotic geometry, modified by Rastall gravity, directly influences observational signatures.

{Phenomenologically, the non-normalizability of the scalar zero mode removes the most dangerous long-range scalar force. The modes found here are massive and have finite lifetimes. For the representative scale $k=10^{-3}\,{\rm eV}$, their frequencies $\nu_n$ are of order $10^{11}\,{\rm Hz}$, their damping times $\tau_n$ are of order $10^{-12}$--$10^{-11}\,{\rm s}$, and their lifetime-limited distance scale $d_n$ is at most of order $10^{-2}\,{\rm m}$. Therefore they are short-distance laboratory-scale modes rather than LISA/PTA-band signals. If interpreted as gravitational-wave signals, they would correspond to scalar-type polarizations rather than the usual tensor polarizations, so their detection would require scalar-polarization sensitivity and a source model for the excitation amplitude. The Rastall parameter then controls the lifetime and the late-time tail shape of any excited scalar signal.}

\acknowledgments
This work was supported by the Scientific Research Projects of Jingchu University of Technology (No.ZD202317), and the Scientific Research Foundation for High-level Talents of Anhui University of Science and Technology (No.2024yjrc164).

\end{document}